\documentclass[twocolumn]{emulateapj}

\usepackage{natbib}
\usepackage{amssymb}
\usepackage{graphicx}

\newcommand{\Msun}     {\mbox{${M}_{\mathord\odot}$}}
\newcommand{\Rsun}     {\mbox{${R}_{\mathord\odot}$}} 
\newcommand{\cxo} {{\em Chandra X-Ray Observatory}}
\newcommand{\ch} {{\em Chandra}}

\begin{document}
\title{The X-ray Halo of Cen X-3}
\author{
Thomas W. J. Thompson\altaffilmark{1} \& Richard E. Rothschild\altaffilmark{1}}
\altaffiltext{1}{Center for Astrophysics and Space Sciences, University of California, San Diego, La Jolla, CA 92093; email: tthompson@physics.ucsd.edu }

\begin{abstract}
Using two {\it Chandra} observations we have derived estimates of the
dust distribution and distance to the eclipsing high mass X-ray binary
(HMXB) Cen X-3 using the energy-resolved dust-scattered X-ray halo. By
comparing the observed X-ray halos in 200 eV bands from 2--5 keV to
the halo profiles predicted by the Weingartner \& Draine interstellar
grain model, we find that the vast majority ($\approx$70\%) of the
dust along the line of sight to the system is located within about 300
pc of the Sun, although the halo measurements are insensitive to dust
very close to the source. One of the \ch~observations occurred during
an egress from eclipse as the pulsar emerged from behind the
mass-donating primary. By comparing model halo light curves during
this transition to the halo measurements, a source distance of $5.7
\pm 1.5$ kpc (68\% confidence level) is estimated, although we find
this result depends on the distribution of dust on very small
scales. Nevertheless, this value is marginally inconsistent with the
commonly accepted distance to Cen X-3 of 8 kpc. We also find that the
energy scaling of the scattering optical depth predicted by the
Weingartner \& Draine interstellar grain model does not accurately
represent the results determined by X-ray halo studies of Cen
X-3. Relative to the model, there appears to be less scattering at low
energies or more scattering at high energies in Cen X-3.

\end{abstract}
 
\section{Introduction}
X-ray halos, which appear as diffuse emission surrounding X-ray
sources, are created by small-angle scattering of soft X-ray photons
from dust grains in the interstellar medium (ISM). Their study can
provide information on interstellar grain properties (density,
morphology, composition) and on the spatial distribution along the
line of sight. Given a dust distribution, variability in the X-ray
halo can be used to geometrically derive estimates of the distances to
X-ray sources based on the time delays of the photons scattered along
the line of sight \citep{ts73}. This method has been applied in a
number of cases \cite[e.g.,][]{predehl00,me06,audley06,xiang07}. The
goal of this work is to estimate the dust distribution and distance to
the eclipsing HMXB Cen X-3 using measurements of its dust-scattered
halo with the \cxo.

The theory of X-ray scattering and the production of X-ray halos has
been described by a number of authors (e.g., Mathis \& Lee 1991, Smith
\& Dwek 1998, Draine 2003). Various interstellar grain models have
been proposed. Mathis et al. (1977) developed a model composed of
silicate and graphite grains with a size distribution of $n(a) \propto
a^{-3.5}$, which reproduces the observed extinction of
starlight. Weingartner \& Drain (2001, hereafter WD01) produced a
grain model that additionally accounts for the diffuse infrared and
microwave emission emission from the ISM by including sufficient small
carbonaceous grains and polycyclic aromatic hydrocarbons
(PAHs). \cite{zubko04} created a series of models by simultaneously
fitting the extinction, infrared emission, and elemental abundance
constraints, by including, for example, amorphous carbon particles,
organic refractory material, water ice, and voids.

One of the many challenges in developing a viable model is that the
characteristics of the dust may vary in different Galactic locations
due to different local evolutionary histories, ISM phases, and
abundances of elements. Given such difficulties, a
preeminent interstellar grain model has not been established. For our
purposes, we use the WD01 grain model, although we allow the energy
dependence of the scattering cross section to be a free parameter.

The basic quantity that determines the characteristics of X-ray halos is
the differential scattering cross section $d\sigma/d \Omega$. This can
be calculated using the exact Mie solution or the simpler
Rayleigh-Gans approximation. For the WD01 grain model, the
differential scattering cross section as a function of scattering
angle $\phi$ can be approximated by the simple analytic form
\begin{equation}
\frac{d\sigma}{d\Omega} \approx \frac{\sigma_{\rm sca}}{\pi \phi^{2}_{{\rm s},50}} \left[1+\left(\frac{\phi}{\phi_{{\rm s},50}}\right)^{2}\right]^{-2}
\end{equation}
where $\phi_{{\rm s},50} \approx 360\arcsec~({\rm keV}/E)$ is the
median scattering angle as a function energy (Draine 2003). It is
important to note that the following results are probably somewhat
dependent on this chosen form of the differential scattering cross
section.

Cen X-3 is one of the brightest accreting X-ray pulsars, and it is one
of the six pulsars in which the observation of eclipses has permitted
the determination of all orbital and stellar parameters. The rotation
period of the pulsar is 4.8 s, and the binary orbital period is 2.1
days. The inclination of the orbital plane and the mass of the neutron
star have been estimated to be $i \sim 70^{\circ}$ and $M = 1.21 \pm
0.21$ \Msun~(Ash et al. 1999). The optical counterpart of Cen X-3 has
been identified as an early type star of radius $R \sim 12$ \Rsun~and
mass $M \sim 20$ \Msun~\citep{rapjoss83}. Accretion in the system
probably occurs via an accretion disk (Bildsten 1997; Takeshima et
al. 1992), although the mass transfer mechanism probably includes an
X-ray excited wind \citep{ds93}.

Although well-studied, the distance to Cen X-3 remains
uncertain. Using a B0 II stellar classification model, an 8 kpc
distance to Cen X-3 was obtained \citep{krz74}, however, subsequent work
indicated that the supergiant primary was of type O6–-8 III (Hutchings
et al. 1979). \cite{day91} observed soft emission
during two eclipses using {\em EXOSAT} and attributed it to the
dust-scattered halo, from which they derived a distance of $5.4 \pm
0.3$ kpc. Discrepancies of $\sim$2.5 kpc in distance estimates to Cen
X-3 lead to nearly a factor of two difference in the inferred optical
and X-ray luminosities. 

 
Cen X-3 ($l=292\fdg09, b=0\fdg34$) is particularly suited for X-ray
halo studies. The source is usually very bright, and the interstellar
hydrogen column density along the line of sight is $1.20 \times
10^{22}$ cm$^{-2}$ \citep{dl90}, implying a large enough optical depth
to result in an appreciable halo \citep{ps95}, yet not so large as to
include a substantial contribution due to multiply scattered
photons. Moreover, the eclipsing nature of the binary system is
advantageous because it should produce the largest fraction of
variability in the X-ray halo, which makes a geometric distance
measurement more direct.

Section 2 presents the general equations that characterize the shape
and variability of X-ray halos. For our study of the X-ray halo of Cen
X-3, two \ch~High Energy Transmission Grating (HETG) observations are
used; one observation took place outside of eclipse ($\Phi_{\rm orb}
\sim 0.4-0.6$) when there was little variability in the flux
(hereafter, called the ``plateau'' phase of the orbit based on the
shape of the light curve), and the other occurred during an egress
from eclipse (\S~\ref{obs}). Because accurate halo measurements
critically depend on careful subtraction of the instrumental point
spread function (PSF), the shape and normalization of the PSF is
parametrized using all \ch/HETG observations of the unabsorbed sources
Her X-1 and PKS 2155-304. These results are presented in the appendix
to the paper. The method for deriving the dust distribution to Cen X-3
is discussed in \S~\ref{distributionsect}. With a dust distribution
established, the large amplitude change in flux during eclipse egress
is used to constrain the distance to the system
(\S~\ref{distancesect}). A comparison of the inferred dust
distribution to observations of CO emission, star counts, and
interstellar extinction in the direction of Cen X-3 is presented in
\S~\ref{compare}. We discuss the implications of our results in
\S~5, and we provide a brief summary in \S~\ref{summary}.

\section{X-ray Halos} \label{halosect}
Following Draine \& Tan (2003; hereafter DT03), for a steady source,
the intensity of single-scattered photons arriving at halo angle
$\theta$ is given by
\begin{equation}
I_{1}(\theta)=F_{\rm X} \tau_{\rm sca} \tilde{I}_{1}(\theta),
\end{equation}
where $F_{\rm X}$ and $\tau_{\rm sca}$ are the flux and scattering
optical depth, respectively, where
\begin{equation}
\tilde{I}_{1}(\theta)\approx \int_{0}^{1} \frac{\tilde{\rho}(x)}{(1-x)^2} \frac{d \sigma}{d \Omega}dx,
\end{equation}
$\tilde{\rho}(x)$ is the dimensionless dust density at fractional
distance $x$ to the source, the scattering angle and the halo angle
are related through $\phi \approx \theta/(1-x)$, and the normalization
is chosen such that $\int \tilde{I}_{1}(\theta) 2 \pi \theta d \theta
=1$. For a variable source, a photon appearing at halo angle $\theta$
after scattering from a dust grain at a fractional distance $x$ to a
source $D$ kpc away will be delayed with respect to the central source
by
\begin{equation}
\delta t=1.21 \frac{D}{\rm kpc} \left(\frac{\theta}{\rm arcsec}\right)^2 \frac{x}{1-x}~{\rm s}.
\end{equation}
In this case, $F_{\rm X}$ must be moved inside the integral (eq. 3)
because the halo flux at time $t$ is proportional the flux of
the central point source at the ``retarded time'' $t^{\prime} = t -
\delta t$, i.e., $I_{1}(t,\theta) \propto F_{\rm X}(t-\delta
t(x,\theta))$.

It is clear that to accurately determine the proportionality between
the source flux and the halo, one must know the history of the source
flux for a sufficiently long time period preceding the observation. In
most cases this is not practical, so typically one assumes that any
change in the source flux took place sufficiently long ago for the
halo to have completely responded to it (as we do when estimating the
dust distribution, \S~\ref{distributionsect}), or one models the
history of the source flux using reasonable estimates (as we do when
constraining the source distance, \S~\ref{distancesect}).

The intensity of the X-ray halo due to photons scattering two or more
times ($I_{2}$, $I_{3}$, \ldots) is straightforward for a steady source
given an assumed dust distribution, and we refer the reader to DT03
for the appropriate recursion formula. However, accounting for the
time delays for multiply-scattered photons is extremely
cumbersome. Fortunately, the intensity of the halo due to photons
scattered $n$-times, relative to the observed source flux, is
$\tau_{\rm sca}^{n}/n!$ \citep{ml91}, so the halo intensity for
multiply-scattered photons decreases rapidly for reasonably small
optical depths ($\tau_{\rm sca} \la 2$). Usually, the first- and
second-order scatterings are sufficient to model the total halo.

\section{Observations \& Data Extraction} \label{obs}
\begin{figure*}
\centerline{\includegraphics[width=7.0in]{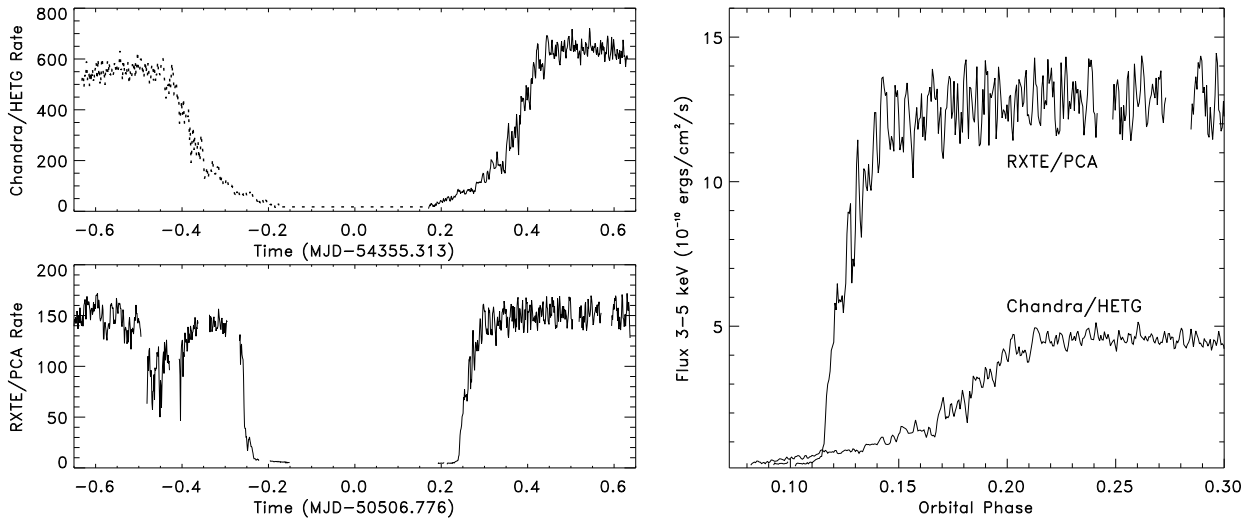}} 
\caption{\footnotesize Comparison of the eclipses and flux states in
  count rates ({\it left panels}) and fluxes ({\it right panel}) from
  3--5 keV between the {\it Chandra} eclipse egress observation (ObsID
  7511) and an {\it RXTE}/PCA observation. No data were obtained for
  the {\it Chandra} eclipse ingress ({\it dotted curve}). The ingress
  is modelled as the mirror image of the egress (multiplied by 0.87,
  see \S~\ref{distancesect}).\label{egresses}}
\end{figure*}

In this paper, we use two {\it Chandra}/HETG observations of Cen X-3
(ObsIDs 1943 \& 7511), and several \ch/HETG observations of PKS
2155-304 and Her X-1 to model the PSF (see the appendix). Although the
use of the grating reduces the zeroth-order image count rate by about
a factor of 3, the dispersed spectrum provides accurate measurements
of the source flux and spectrum, which is necessary to accurately
model the characteristics of the X-ray halo and the PSF. One of the
observations (ObsID 1943) began at MJD 51908.01 and lasted for 45 ks,
corresponding to $\Phi_{\rm orb} \approx 0.38-0.63$, where
determination of the orbital phases was accomplished using the
mid-eclipse ephemeris from Burderi et al. (2000), and the orbital
period and evolution $[\mbox{\.{\em P}}_{\rm orb}/P_{\rm orb} =-(1.738
\pm 0.004) \times 10^{-6}$ yr$^{-1}]$ from Nagase et al. (1992).  A
dip in the flux was observed at about $\Phi_{\rm orb} \approx 0.57$,
so we ignored data after this time to simplify analysis by removing
any time dependence (eq. 4). X-ray halo measurements from this
observation were used to estimate the dust distribution along the
line of sight. The other observation of Cen X-3 (ObsID 7511), which
resulted from a successful observing proposal with the primary goal of
geometrically estimating the distance to the system, took place during
an egress from eclipse. The 40 ks observation began at MJD 54355.46
($\Phi_{\rm orb} \approx 0.08-0.30$). The preceding mid-eclipse epoch
was calculated to be MJD $54355.313 \pm 0.007$. Figure \ref{egresses}
(top-left panel, solid curve) shows the \ch/HETG light curve during
the eclipse egress observation. We also present an {\it RXTE}/PCA
light curve for comparison. Clearly, the \ch~observation occurred
during a relatively lower flux state, leading to a more gradual
increase in flux coming out of eclipse as compared the PCA
observation.  \cite{pt83} discussed the aperiodic 120--165 day
timescale between the different flux states, and \cite{clark88}
studied how the stellar wind and the shape of the egress light curves
are affected by the different states.

Data analysis was performed using the standard tools of the Chandra
Interactive Analysis of Observations (CIAO) software version 4.0 and
Calibration Database (CALDB) version 3.4.3. A single first-order
Medium Energy Grating (MEG) and High Energy Grating (HEG) spectrum was
extracted for the plateau phase observation (ObsID 1943). For the
eclipse egress observation (ObsID 7511), ten 3968 s first-order MEG/HEG
spectra were extracted to model the evolution of the point source
spectrum. Spectral models were fit to each spectrum using XSPEC
version 11.3, and fluxes (in units of [photons/cm$^{2}$/s]) were
measured in 15 energy bands of 200 eV width from 2--5 keV. Finer time
resolution on the flux evolution was obtained by comparing the average
HETG count rates in each time range and energy band to the relative
count rates in smaller 100 s time bins.

Separate X-ray halo images and exposure maps were extracted in 200 eV
bands from 2--5 keV, from which we obtained exposure-corrected surface
brightness distributions using 40 logarithmically-spaced annular
regions surrounding the source. For the plateau phase observation, a
single exposure map was created for each energy band. On the other
hand, accurate measurements of the evolution of the halo flux in each
energy band during the eclipse egress observation required the use of
separate exposure maps (due to the dither of the telescope) using 500
s bins (1200 exposure maps in total). Due to issues with the
diffracted HETG photons confusing the background analysis (HETG
photons create a diffuse ``transfer swath'' analogous to the
``transfer streak'' in the zeroth-order image), we did not use halo
angles greater than 110\arcsec. The surface brightness distributions
were divided by the flux measurements in each band to produce images
in units of flux fraction per square arcsecond. Halo angles smaller
than about 3\arcsec~are affected by pile-up and ignored. To minimize
concerns due to multiply-scattered photons, we excluded energies below
2 keV in each observation.

\begin{figure*}
\centerline{\includegraphics[width=7.in]{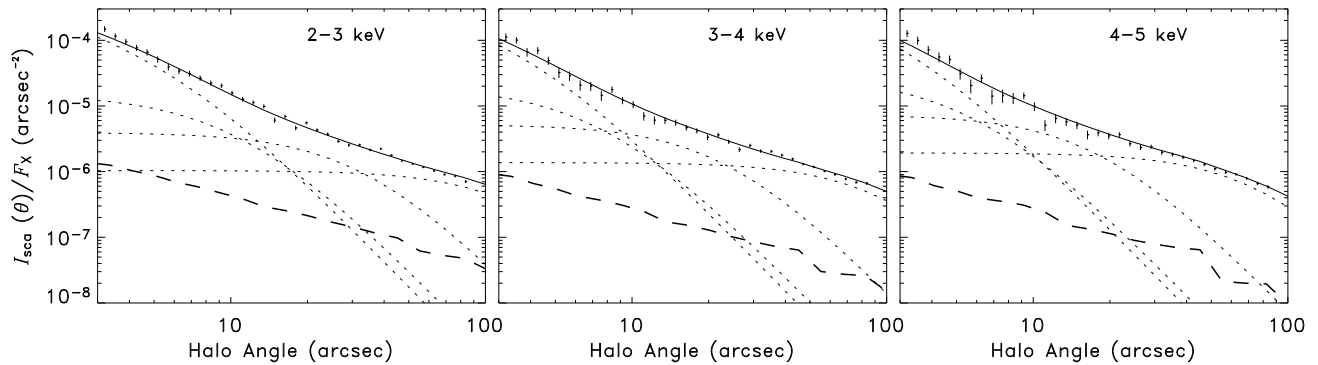}} 
\caption{\footnotesize Cen X-3 halo measurements and best-fit model
  halos (\S~\ref{distributionsect}) shown in three bands from 2--5 keV
  in units of source flux per square arcsecond. Note that the actual
  fits to the halos used 200 eV bands. Larger 1 keV bins are displayed
  to facilitate the visual comparison between the models and the
  data. The {\it dotted} curves show the contribution to the halo from
  the individual single-scattered ($\Delta x =0.05$) clouds (see
  Fig. \ref{sumhalo} for the cloud locations), the {\it dashed} curves
  show the halo contribution from photons scattered twice, and the
  {\it solid} curves show the sum of the single- and double-scattered
  halos ($\chi^{2}_{\nu}=1.05$, 502 dof). Third- and higher-order
  scatterings are nearly negligible and are not included in the
  model. \label{fitdust}}
\end{figure*}

In the appendix, we parametrize the PSF as the fraction of the point
source flux comprising the PSF as a function of halo angle. Although
the observed surface brightness distributions represent the
convolution of the PSF and the X-ray halo, in our analysis we treat it
as the sum. This is acceptable because we restrict our investigation
to angular scales that are much larger than the
$\sim$0\farcs5~resolution of the telescope mirrors. The PSF
contribution in each energy band is simply subtracted from the
observed surface brightness distribution. The resulting X-ray halos,
rebinned in 1 keV bands, are shown in Figure \ref{fitdust}. The fitted
halo components are discussed below.

\section{Analysis \& Results}
\subsection{Estimating the Dust Distribution} \label{distributionsect}
\begin{figure}
\centerline{\includegraphics[width=3.5in]{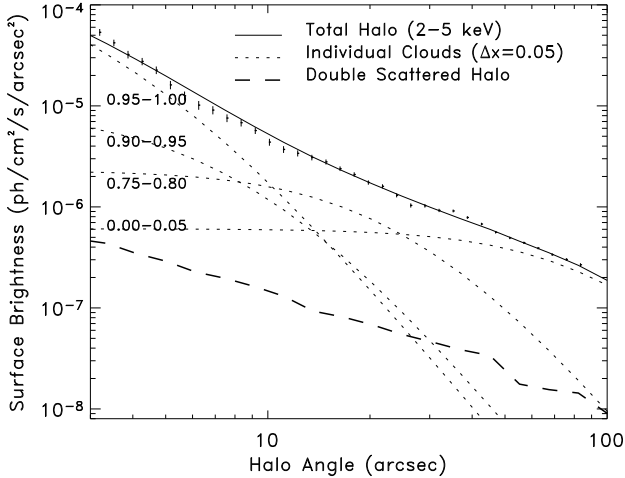}} 
\caption{\footnotesize Surface brightness distribution (with PSF
  subtracted) from 2--5 keV. As in Fig. \ref{fitdust}, the {\it
    dotted} curves show the contribution to the halo from the
  individual ($\Delta x =0.05$) clouds and the {\it dashed} curves
  show the halo contribution from photons scattered twice. The
  location of the dust for each single-scattered halo component is
  labelled at left (also see Fig. \ref{distribution}).\label{sumhalo}}
\end{figure}

In order to estimate the dust distribution to Cen X-3, we used the
plateau phase \ch~observation. We {\it assumed} the halo had
sufficient time to respond to the eclipse egress, which based on the
orbital ephemeris occurred $\sim$30 ks earlier. Once we established a
dust distribution, we checked whether or not the assumption was
acceptable. First, we created separate single-scattered X-ray halo
profiles for twenty step-function dust distributions of width $\Delta
x =0.05$ (hereafter referred to as ``clouds''), spanning the entire
line of sight, i.e., 0.00 to 0.05, 0.05 to 0.10, and so on, for each
energy band. By summing the twenty profiles with equal weighting, one
could obtain the single-scattered X-ray halo for dust distributed
uniformly along the line of sight. The normalization of each cloud
($a_{x}$), which is proportional to the contribution to the total
scattering optical depth (due to the normalization $\int
\tilde{I}_{1}(\theta) 2 \pi \theta d \theta =1$), was allowed to vary
freely. The initial fit function at each energy is
\begin{equation}
\frac{I_{\rm sca}(\theta)}{F_{\rm X}}=\sum\limits_{x=0}^{19} a_{x} \tilde{I}_{1}^{x}(\theta),
\end{equation}
where
\begin{equation}
\tilde{I}_{1}^{x}(\theta)\approx \int_{0.05x}^{0.05(x+1)} \frac{1}{(1-x)^2} \frac{d \sigma}{d \Omega}dx.
\end{equation}
Note that the superscript in $\tilde{I}_{1}^{x}$ is a label that
indicates a position along the line of sight and is not an
exponent. Using this method, the dimensionless dust density and
scattering optical depth are simply
\begin{equation}
\tilde{\rho}(x)=\frac{a_{x}}{\sum a_{x}},~\sum a_{x}=\tau_{\rm sca},
\end{equation}
respectively.  The halos for each energy band were fit
simultaneously. For the initial fit, we used the energy scaling of the
scattering optical depth from the WD01 interstellar grain model
($R_{V} = 3.1$); however, we found it to underpredict the halo flux at
higher energies (or overpredict the flux at lower energies). We
therefore allowed the energy scaling to also be a free parameter,
although we required that it be a smooth function, i.e., $\tau_{\rm
  sca}(E) \propto (E-E_{0})^{-\beta}$, where $E_{0}$ and $\beta$ are
fit parameters. Nevertheless, the angular dependence of the scattering
at a particular energy still follows the functional form predicted by
the WD01 model (see eq. [1]).  The best-fit normalization of each
cloud provided a preliminary dust distribution. Given the preliminary
model, the X-ray halo for double-scattered photons was calculated
($I_{2}(\theta)$), included in the model as an additional term in
eq. (5), and the normalization for each cloud was refit. This process
was repeated in an iterative manner to coverage to the dust
distribution whose single- and double-scattered X-ray halos accurately
described the Cen X-3 halo. Third- and higher-order scatterings were
ignored, but their contribution to the halo is negligible. The Cen X-3
X-ray halos in units of source flux and the best-fit halo models are
presented in Figure \ref{fitdust}, and the integrated (2--5 keV)
surface brightness distribution is shown in Figure
\ref{sumhalo}. Evidently, the 3\arcsec--100\arcsec~halo can be modeled
using only four separate dust clouds (Fig. \ref{sumhalo}, dotted
curves).

\begin{figure}
\centerline{\includegraphics[width=3.5in]{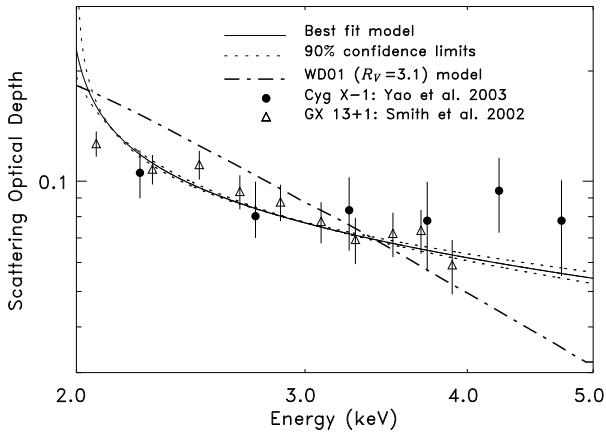}} 
\caption{\footnotesize Best-fit scattering optical depth to Cen X-3 as
  a function of energy ({\it solid} curve). The empirical best-fit
  curve is described by $\tau_{\rm sca}(E) \propto
  (E-E_{0})^{-\beta}$, where $E_{0} \sim 1.96$ and $\beta \sim
  0.32$. The optical depth predicted by the WD01 interstellar grain
  model ($R_{V}=3.1$) is shown as a dash-dotted curve. The
  normalization for the WD01 model was chosen so that the areas under
  each curve are the same. Also included are Cyg X-1 data from
  \cite{yao03} ({\it filled circles}) and GX 13+1 data from
  \cite{smith02} ({\it triangles}) (see discussion). The
  normalizations for the Cyg X-1 and GX 13+1 data have been modified
  to more closely align the points with the empirical curve for Cen
  X-3. The data points from \cite{yao03} have been corrected assuming
  uniform dust, and the data points from \cite{smith02} have been
  corrected assuming the dust density is proportional to distance. See
  Draine (2003) for further details on the correction method.
\label{tau}}
\end{figure}

The energy dependence of the scattering optical depth is presented in
Figure \ref{tau}, showing the relatively brighter halo at higher
energies as compared to the WD01 interstellar grain model predictions
(see discussion). The Cyg X-1 and GX 13+1 points are the same data as
presented in Fig. 11 of Draine (2003), who compared the optical depth
calculated from the WD01 model to observations. Although Draine (2003)
extended his comparison to other sources, only Cyg X-1 and GX 13+1
span the 2--5 keV energy range that we investigate for Cen X-3.


\begin{figure}
\centerline{\includegraphics[width=3.5in]{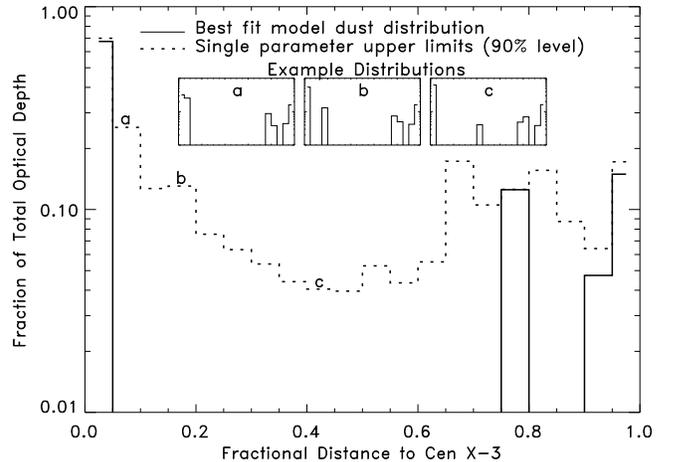}} 
\caption{\footnotesize Best-fit model dust distribution to Cen X-3
  ({\it solid histogram}). The majority of the dust is located near
  the Sun ($0.00 < x < 0.05$), and the other primary dust locations
  are at $x>0.75$. Although the best-fit distribution accurately
  models the data ($\chi^{2}_{\nu}=1.05$, 502 dof), the size of any
  individual dust cloud is not well-constrained ({\it dotted
    histogram}). The three small panels labeled $a$, $b$, and $c$
  (corresponding to the upper limits shown in the main panel) show
  examples of dust distributions that are also consistent with the data at
  the 90\% confidence level. \label{distribution}}
\end{figure}

One of the primary goals of this work is to geometrically constrain
the distance to Cen X-3, therefore it is important to quantify the
uncertainty in the dust distribution because this directly affects the
size of the time delays of photons scattered along the line of
sight. By fixing the double-scattered halo model to the one
corresponding to the best-fit dust distribution, the single-parameter
90\% confidence uncertainties in the contribution from each of the
twenty clouds was calculated. In each case the lower limit on the
cloud size was zero, i.e. no dust, simply because the clouds to either
side of the cloud in question could produce a surface brightness
distribution with a similar shape. On the other hand, the upper limits
to the size of each cloud are well-determined. Figure
\ref{distribution} shows the best-fit dust distribution towards Cen
X-3 and the single-parameter upper limits for the size of each
cloud. Three examples of different dust distributions that are also
consistent with X-ray halo data are presented. The majority of the
dust ($\approx$70\%) in each case is located at small fractional
distances to Cen X-3. Note, however, that we do not have reliable data
for halo angles less than about 3\arcsec, so we are mostly insensitive
to dust that produces centrally-peaked halos, i.e., dust very near the
source ($x>0.99$). Therefore, it is more accurate to consider the
scattering optical depth in Fig. \ref{tau} as a lower limit.

It could be argued that lack of knowledge of the dust very near the
source poses a problem for our distance estimate. This is not the
case: For an assumed distance of 8 kpc, the $\sim$0\farcs5 angular
resolution of the \ch~mirrors corresponds to approximately 4000 AU,
which, based on the orbital parameters of the binary system, is more
than 4 orders of magnitude larger than the semi-major axis of the
orbit. Therefore, the point source flux already includes all of the
effects due to scattering in the vicinity of the binary system.

\begin{figure} 
\centerline{\includegraphics[width=3.5in]{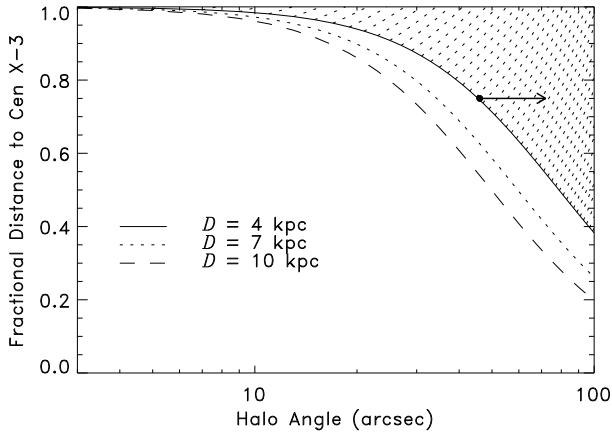}} 
\caption{\footnotesize Phase space boundaries ($x$ vs. $\theta$)
  showing causally-connected regions for delay times less than 30
  ks. The hatched regions of phase space have yet to respond to
  changes in the source flux 30 ks earlier (see eq. 4). For example,
  assuming a source distance of 4 kpc and dust at $x=0.75$, for
  $\theta > 46\arcsec$ the halo has not yet responded to changes in
  the source flux 30 ks earlier (see point and arrow). \label{boundary}}
\end{figure}

Considering the best-fit dust distribution, it is clear that the
assumption that the halo during the plateau phase observation had
responded to the preceding eclipse egress is not valid for all halo
angles. As pointed out above, an estimate of the time Cen X-3 has been
out of eclipse at the beginning of the observation is about 30 ks. The
causally-connected regions of $x$ vs. $\theta$ phase space for delay
times less than 30 ks are shown in Figure \ref{boundary} for three
source distances, indicating that the assumption is invalid for large
$x$ and $\theta$. Fortunately, the vast majority of the dust along the
line of sight is located at $x<0.05$ which presents no
inconsistency, and the two clouds closest to the source ($0.90<x$)
dominate the halo at $\theta\la10\arcsec$ (see Fig. \ref{sumhalo}),
where the halo angles are small enough to have responded to the
previous eclipse egress. The only potential problem concerns the
$x=[0.75,0.80]$ dust cloud, which for $\theta\ga30\arcsec$ have not
had sufficient time to respond to source flux variability. However,
the dust cloud at $x<0.05$ also begins to dominate for sufficiently
large halo angles, so we conclude that the assumption is reasonably
valid.

\subsection{Constraining the Source Distance} \label{distancesect}
\begin{figure}
\centerline{\includegraphics[width=3.7in]{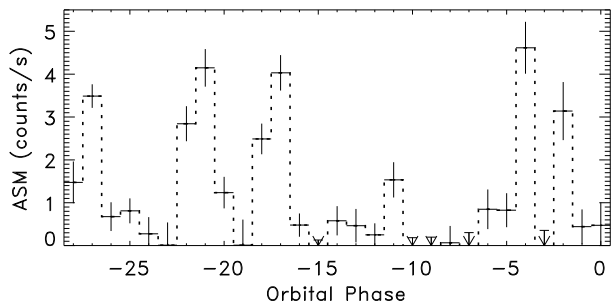}} 
\caption{\footnotesize {\it RXTE}/ASM light curve of Cen X-3 for the 60 days (corresponding to $x=0.98$ for $D=8$ kpc and $\theta=100\arcsec$) preceding the {\it Chandra} observation during eclipse egress. The long-term Cen X-3 light curve was modeled using the {\it Chandra} fluxes scaled by the ASM count rate. Orbital phase 0.0 corresponds to MJD 54355.31.\label{asm}}
\end{figure}
Because our goal is to use the Cen X-3 halo during the eclipse egress
to constrain the distance to the system, it is necessary to model the
source flux history for a sufficiently long time period preceding the
observation (\S~\ref{halosect}). We chose to model the flux history by
first assuming the shape of the light curve during the eclipse ingress
in each energy band is same as the shape during egress, but scaled to
87\% of flux after egress (Fig. \ref{egresses}, top-left panel, dotted
curve). It is widely known that the Cen X-3 flux decreases after
$\Phi_{\rm orb} \sim 0.5$, and the scaling factor represents the
average 2--10 keV flux change that was observed during two consecutive
binary orbits by \cite{suchy08}. So-called ``pre-eclipse dips'' are
also commonly seen in Cen X-3 light curves (Fig. \ref{egresses},
lower-left panel), however, we did not include these features in our
model. Second, for the 60 days prior to the observation we modeled the
flux history by multiplying the \ch~light curve (and modeled ingress)
by the relative {\it RXTE}/ASM count rate at the time (linearly
interpolating the ASM rates at intervening times). The ASM light curve
for Cen X-3 is shown in Figure \ref{asm}. 

Admittedly, the modeled flux history is rather approximate. Not only
have we ignored the pre-eclipse dips, but the assumption that the
shape of the Cen X-3 light curve is independent of the absolute flux
(the relative ASM count rates span $\sim$$0.05\times$ to $5\times$ of
the flux during \ch~observation; Fig. \ref{asm}) is incorrect (see
Fig. \ref{egresses}, right panel). Yet most of the flux at
sufficiently large halo angles is due to dust located within 300 pc of
the Sun. The corresponding delay times are small, and the precise
history of the source light curve can be measured directly from the
\ch~observation. Therefore, any inaccuracies in the modeled flux
history will have a very limited effect, so a crude model may be
assumed sufficient. On the other hand, at smaller angles the halo can
be attributed to dust at larger $x$, making any inaccuracies in the
flux history model more problematic. For this reason, we ultimately
chose not to investigate halo angles smaller than
$\sim$20\arcsec. Separate 200 eV PSF-subtracted surface brightness
distributions were extracted, using time bins of 1.5 ks. The energy
bands were summed together, and halo light curves were created for
four annular regions: $[22\farcs8,33\farcs5]$,
$[33\farcs5,49\farcs2]$, $[49\farcs2,72\farcs2]$, \&
$[72\farcs2,106\farcs0]$.
\begin{figure}
\centerline{\includegraphics[width=3.5in]{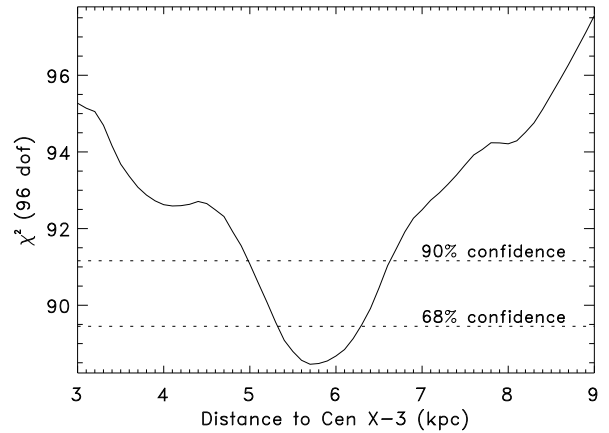}} 
\caption{\footnotesize Fit statistic ($\chi^{2}$) versus distance to
  Cen X-3. At the 68\% (90\%) confidence level, the distance is
  5.7$^{+0.5}_{-0.3}$ ($^{+0.9}_{-0.7}$) kpc. The quality of the fit
  at the best-fit distance is $\chi^{2}/{\rm dof} =87.0/96=0.91$. A
  more appropriate method for error determination is discussed
  below.\label{dist}}
\end{figure}

\begin{figure*}
\centerline{\includegraphics[width=7in]{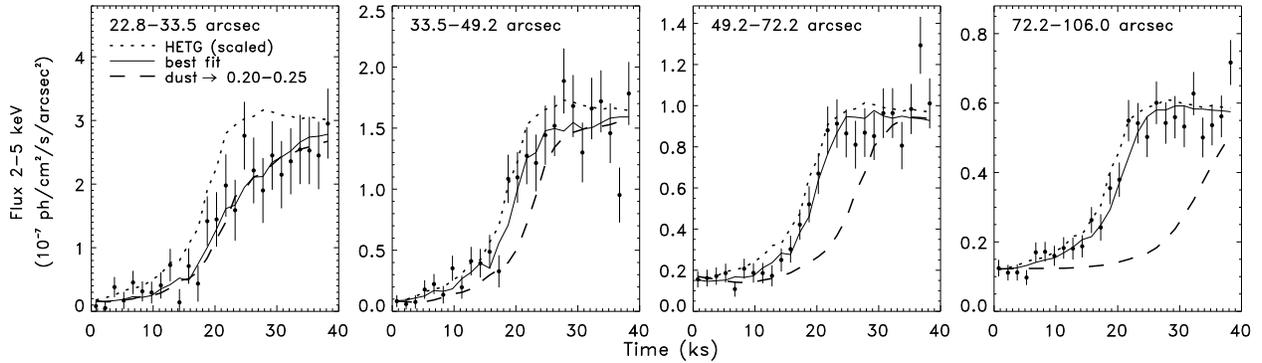}} 
\caption{\footnotesize Eclipse egress light curves from 2--5 keV for
  four annular regions (labeled in each panel). The model light curves
  for $D=5.7$ kpc are shown as {\it solid} curves. The {\it dotted}
  curves show the HETG light curve, and the {\it dashed} curve shows
  what the predicted light curves would be if the prominent dust cloud
  was shifted from $x=[0.00,0.05]$ to $x=[0.20,0.25]$.\label{fits}}
\end{figure*}
Using the model flux history, we created model halo light curves for
various energies, halo angles, and source distances, using the
best-fit dust distribution as well as the three example dust
distributions that are also consistent with the X-ray halo data. The
final 2--5 keV single-scattered model light curves for the four
annular regions are then derived by summing the energy bands and
integrating over the annular regions, {\it viz.}
\begin{equation}
I_{1}(t,\Delta \theta)=\sum\limits_{i=0}^{14}\frac{1}{\pi (\theta_{2}^{2}-\theta_{1}^{2})}\int_{\theta_{1}}^{\theta_{2}} 2 \pi \theta I_{1}^{i}(t,\theta)d \theta,
\end{equation}
where the superscript $i$ is a label that indicates the center of the
200 eV energy band at $(2.1+0.2 i)$ keV, and $\theta_{1}$ and
$\theta_{2}$ are the boundaries of each annular region. 

Although the single-scattered halo light curves can be easily modeled,
we must also account for multiply-scattered photons, as well as our
ignorance of the specific form of the dust distribution
(Fig. \ref{distribution}). Calculating the time dependence of
multiply-scattered photons is computationally cumbersome
(DT03). Fortunately, these photons comprise $\sim$10\% of the total
halo flux over the angles of interest (Fig. \ref{sumhalo}), so a
rather simple treatment should be sufficient. Moreover, because the
dust is concentrated at small $x$, the response of the halo to
multiply-scattered photons should occur almost as rapidly as the
response to single-scattered photons. Thus, to roughly account for
multiply-scattered photons, we added 10\% (the approximate value of
$I_{2}/I_{1}$ over the angles of interest in the plateau phase
observation) to the normalization of the single-scattered model light
curves. We also allowed the normalization to vary by 10\% when
obtaining the fits.

To address the uncertainties in the dust distribution, we calculated
model light curves for the best-fit distribution and for the three
example distributions shown in Fig. \ref{distribution}. Although the
general characteristics of the light curves (shape and normalization)
for the different dust distributions were similar, subtle differences
were present. The size of the differences in the light curves were
about 7\% $\pm$ 3\% (relative to the model light curve for the
best-fit distribution) at the beginning of the observation (and for
the most part, independent of distance and angle), and 4\% $\pm$ 1\%
at the end. Clearly, these measured differences only apply to the
selected set of four potential dust distributions. In each case,
however, the majority of the dust is located relatively close to the
Sun, so the differences in the light curves from this set should
provide a decent measure of the uncertainty in $I_{1}(t,\Delta
\theta)$ due to the uncertainty in $\tilde{\rho}(x)$. We chose to
account for the uncertainty in the dust distribution by adding
$\sim$4\%--7\% (depending on differences in the model light curves as
a function of time) to the intrinsic uncertainty of the halo light
curves.

\begin{figure}
\centerline{\includegraphics[width=3.5in]{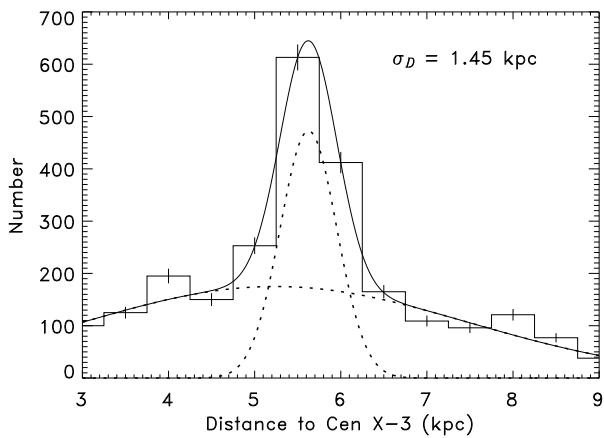}} 
\caption{\footnotesize Histogram of the best-fit distances to Cen X-3
  from 2500 synthetic data sets. The distribution is fitted with the
  sum of two Gaussians. The standard deviation of the distribution is
  1.45 kpc, and the 90\% confidence level lower and upper limits,
  which are determined from the {\it model} of the distribution, are
  2.1 kpc and 8.3 kpc, respectively.\label{derr}}
\end{figure}

Finally, the distance to Cen X-3 was obtained by fitting the model
halo light curves for the four annular regions and for source
distances between 3 kpc and 12 kpc in steps of 0.1 kpc to the observed
light curves. For each distance, the $\chi^{2}$ statistic was used to
determine the quality of the fit. Our results indicate that the
distance to Cen X-3 is $5.7^{+0.5}_{-0.3}$ ($^{+0.9}_{-0.7}$) kpc with
68\% (90\%) confidence (Figure \ref{dist}). Note that we derive a more
reasonable estimate of the distance uncertainty below.

The comparison of the measured halo light curves to the model fits for
a source distance of 5.7 kpc (solid curves) is shown in Figure
\ref{fits}. Also shown is the model light curves if the
$x=[0.00,0.05]$ dust cloud were moved to $x=[0.20,0.25]$. The halo
light curves are therefore clearly consistent with most of the dust
having a local origin.

\begin{figure}
\centerline{\includegraphics[width=3.5in]{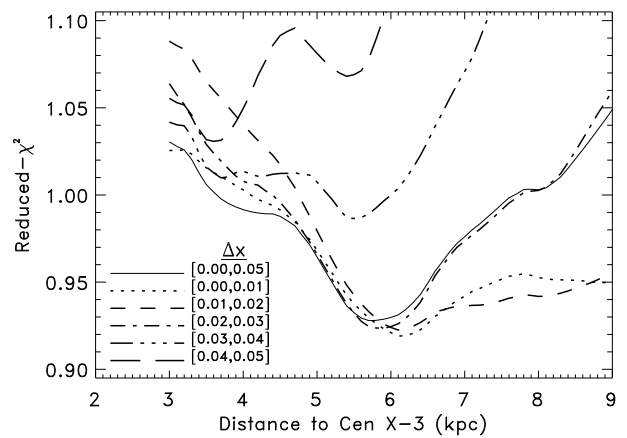}} 
\caption{\footnotesize Similar to Fig. \ref{dist}, but decreasing the
  width of the $x=[0.00,0.05]$ dust cloud to $\Delta x=0.01$ (while
  keeping the total amount of dust in the cloud constant by
  multiplying the normalization by 5) and trying each position within
  $x=[0.00,0.05]$.\label{diffdist}}
\end{figure}


Because the model light curves are not strictly linear in distance,
$\Delta \chi^{2}$ may not be a legitimate estimator of the distance
uncertainty (Fig. \ref{dist}). A more reasonable estimate of the
uncertainty can be obtained using Monte Carlo simulations. We
accomplished this by simulating 2500 synthetic data sets, assuming the
light curve models for $D=5.7$ kpc accurately describe physical
reality, and assuming the measurement uncertainties are normally
distributed. For each realization the best-fit distance is determined
from the $\chi^{2}$ minimum. The resulting probability distribution of
best-fit distances is presented in Figure \ref{derr}, indicating that
the uncertainty in distance based on $\Delta \chi^{2}$ is indeed too
small, and that a more reasonable distance estimate to Cen X-3 is $5.7
\pm 1.5$ $(^{+2.6}_{-3.6})$ kpc with 68\% (90\%) confidence.  The 68\%
error limits were determined simply from the standard deviation of the
distribution, and the 90\% limits were determined by integrating over
the model of the distribution, which is composed of the sum of two
Gaussians.

\begin{figure}
\centerline{\includegraphics[width=3.5in]{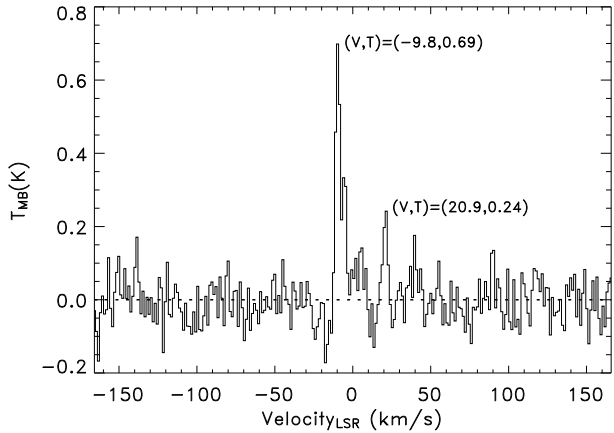}} 
\caption{\footnotesize CO emission observed in the direction of Cen
  X-3 (Dame et al. 2001), with the two most prominent emission
  features labeled. \label{co}}
\end{figure}

Another issue to address is the use of $\Delta x=0.05$ bins for the
dust clouds. The uncertainty in the position of the dust {\it within}
the $x=[0.00,0.05]$ cloud can have a significant effect on the model
halo light curves. To test the effect of various dust positions for
$x<0.05$ we calculated model light curves for the five positions with
$\Delta x=0.01$ in $x<0.05$, maintaining a constant amount of dust in
the cloud by increasing its normalization by a corresponding amount.
For each of the five trial dust positions, the fit statistic was
calculated at each distance (Fig. \ref{diffdist}). The results are
somewhat troubling, because although dust positions at $x<0.04$ do not
greatly affect the implied distance, dust oriented at $x=[0.04,0.05]$
produce two local minima, with the best-fit distance becoming very
small.

\subsection{Comparison of Dust Distribution to CO Emission, Star Counts, and Interstellar Reddening} \label{compare}
\begin{figure}
\centerline{\includegraphics[width=3.5in]{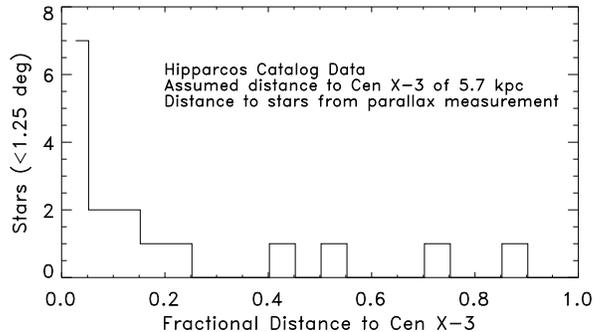}} 
\caption{\footnotesize Histogram of the number of stars within 1\fdg25 of the line of sight to Cen X-3. Data are based on the {\it Hipparcos Catalog}. The fractional distance assumes a source distance of 5.7 kpc.\label{hippar}}
\end{figure}

\begin{figure}
\centerline{\includegraphics[width=3.5in]{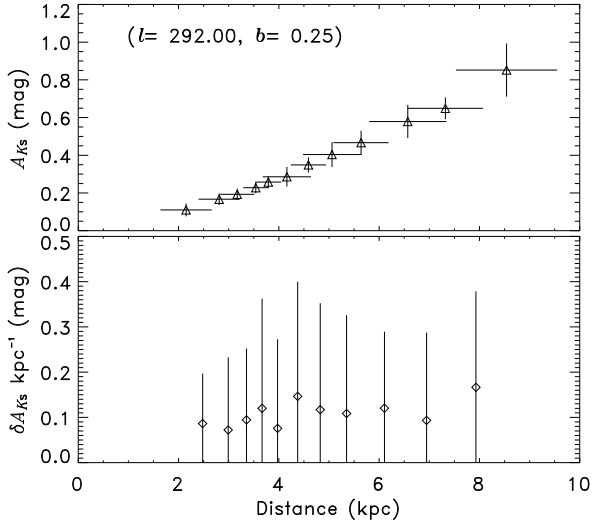}} 
\caption{\footnotesize Interstellar extinction ($A_{K_{\rm s}}$ and $\delta A_{K_{\rm s}}~{\rm kpc}^{-1}$) versus distance in the direction of Cen X-3 (centered 0\fdg13 away from the sightline). Data are from \cite{mar06}.\label{aks}}
\end{figure}

The dust distribution to Cen X-3 can be checked for consistency in a
number of ways. Here we apply three methods. First, using the results
of Dame et al. (2001), we can map the CO emission in the direction of
Cen X-3, which presumably also traces the dust. The most prominent
peak in the emission has a velocity of $-9.8$ km/s (Fig. \ref{co}).
The kinematic distance to the CO responsible for this feature could be
estimated from the rotation curve of the Galaxy, but because molecular
clouds have a cloud-cloud dispersion of 4$-$5 km s$^{-1}$, kinematic
distances for clouds with $|v|<10$ km s$^{-1}$ are not reliable; all
that can be said is that CO emission has a relatively local origin
(T.~M. Dame, priv. comm.). Second, parallax measurements provide the
spatial density of stars along the line of sight to Cen X-3. Using
data from the {\it Hipparcos Catalog}, we created a histogram of the
number of stars within 1\fdg25 of the line of sight (Figure
\ref{hippar}). Of course, the parallax data are subject to certain
selection effects. For one, stars at larger distances are less likely
to be above the flux limit of the survey. On the other hand, a given
solid angle corresponds to a larger volume of space at larger
distances. Nevertheless, assuming star-forming regions are cospatial
with regions of higher dust density, the histogram of star counts
supports that conclusion that most of the dust is located at
relatively small distances from the Sun. Finally, interstellar
reddening data may be able to constrain the dust
distribution. \cite{mar06} modeled the interstellar extinction
distribution in the Galaxy by comparing the infrared colors measured
by the Two Micron All Sky Survey (2MASS; Cutri et al. 2003) to the
colors from the Besan\c{c}on model of the Galaxy \citep{robin03}. The
resulting data for the Cen X-3 sightline, which shows the interstellar
extinction ($A_{K_{\rm s}}$), are presented in Fig. \ref{aks} (top
panel). The constant slope of $A_{K_{\rm s}}$ indicates that the dust
is uniformly distributed between about 2 and 9 kpc. Because there are no
data for $D\la2$ kpc, these interstellar reddening data cannot support
or refute the claim for the local origin of the dust towards Cen
X-3. However, by dividing the difference in extinction between
neighboring bins by the distance between them, one can visualize the
distribution of extinction elements along the sightline (Figure
\ref{aks}, bottom panel) for $D>2$ kpc. The results for this region
of space are consistent with the dust being uniformly distributed,
although we point out that the error bars allow for a slight increase 
in dust density for $D\ga3.5$ kpc, which is coincident with a slight
increase in the upper limits to the size of the dust clouds shown in 
Fig. \ref{distribution} for $x>0.65$. 


\section{Discussion}
In this paper, we have discovered that most of the dust along the line
of sight to Cen X-3 is located relatively nearby the Sun, and that the
distance to the binary system may be smaller than the commonly assumed
distance of 8 kpc. Although our distance estimate is not particularly
robust, we note that the best-fit distance and 8 kpc are only
marginally inconsistent at the 68\% confidence level, and the best-fit
distance is consistent with 8 kpc at the 90\% level.

Despite the uncertainty in the distance to Cen X-3, by all indications
the dust along the line of sight to the system is heavily concentrated
within about 300 pc of the Sun. Presumably, this dust is from the
local Orion spur where the Sun resides. Not only do the X-ray halo
measurements, the CO emission measurements, and the density of stars
along the line of sight support this idea, but the halo light curves
during egress clearly show a rapid response to changes in the point
source flux (Fig. \ref{fits}). In fact, model light curves for a
uniform dust distribution (not shown), or for a distribution where the 
primary dust cloud is $x=[0.20,0.25]$, grossly misrepresent the data.

It is worth examining the difference between the energy
dependence of the scattering optical depth predicted by the WD01
interstellar grain model, and the empirical curve describing Cen X-3
as well as data from Cyg X-1 and GX 13+1 (Fig. \ref{tau}). Clearly,
the shape of the WD01 curve and the Cen X-3 empirical curve differ
substantially. One possible explanation for the discrepancy is our
choice of $R_{V}=3.1$. After all, higher values of $R_{V}$ are
observed in dense clouds, and we have found that most of the dust
along the line of sight is located in a relatively small region of
interstellar space. Yet this is not a viable explanation because the
difference in the scattering optical depth for different values of
$R_{V}$ (over 2--5 keV) is more in the normalization and not the {\it
  shape} of the curve. For a given optical depth at 2 keV, the
increase in optical depth at 5 keV for $R_{V}=4.0$ ($R_{V}=5.5$) is
6--7\% (14--15\%), whereas a $\sim$60--80\% difference is required to
explain the energy dependence of $\tau_{\rm sca}$ in Cen
X-3. Furthermore, the energy scaling of the scattering optical depth
in Cyg X-1 and GX 13+1 more closely resembles the empirical curve for
Cen X-3 than the WD01 model curve. In his comparison, Draine (2003)
focused on the apparent differences between the predicted and measured
values of $\tau_{\rm sca}/A_{V}$ for various X-ray sources, and not on
the shape of the energy dependence. Our results, and those of
\cite{smith02} and \cite{yao03}, suggest that the WD01 model fails to
accurately reflect the energy dependence of the scattering optical
depth. Relative to the WD01 model, there appears to be less scattering
at low energies, or alternatively, more scattering at high energies
(due to uncertainties in the absolute normalization of the scattering
optical depth) in Cyg X-1, GX 13+1, and Cen X-3.


\section{Conclusions} \label{summary}
\begin{enumerate}
\item The vast majority of the dust along the line of sight to Cen X-3
  is located within 300 pc ($x < 0.05$) of the Sun. The most
  likely location of the dust is the local Orion spur. However, we are
  not sensitive to dust located in the vicinity of Cen X-3 because
  data within $\sim$3\arcsec~of the central source are affected by
  pile-up.

\item The geometric distance to Cen X-3 is $5.7 \pm 1.5$ kpc (68\%
  confidence level), although this result is not valid if a large
  fraction of the dust is concentrated within $\Delta x =0.01$.

\item The energy scaling of the scattering optical depth predicted by
  the WD01 interstellar grain model does not accurately represent the
  results determined by X-ray halo studies of Cen X-3. Relative to the
  WD01 model, there appears to be less scattering at low energies or
  more scattering at high energies in Cyg X-1, GX 13+1, and Cen X-3.
\end{enumerate}

\acknowledgements We thank T.~M. Dame for providing the CO emission
data that was used to create Fig. \ref{co}. We also acknowledge that
D.  Marshall provided the data for Fig. \ref{aks}. We also thank the
anonymous referee for helpful suggestions that
improved on the first draft of the paper. This work was supported by
NASA contract NAS5-30720 and {\it Chandra} grant GO7-8046X.

\appendix{} 
\begin{center}
{\it Chandra}/HETG {\sc Point Spread Function}
\end{center}

The excellent angular resolution of the \ch~High Resolution Mirror
Assembly (HRMA) and the good spectral resolution of the Advanced CCD
Imaging Spectrometer (ACIS) provide the best opportunity to date for
the study of X-ray halos. However, careful subtraction of the Point
Spread Function (PSF) is still required to produced accurate
results. The raw images of X-ray sources are the convolution of the
X-ray halo surface brightness distribution and the PSF of the mirror
assembly, modified by the instrumental response of the CCDs, the
reduction of effective area at larger off-axis angles due to
vignetting, and the dithering of the telescope throughout the
observation. Exposure maps account for all of these effects, with the
exception of the PSF. At small halo angles ($\la 5 \arcsec$)
MARX\footnote{http://space.mit.edu/ASC/MARX/} simulations can be used
to model the PSF, but at larger halo angles the results of MARX
simulations appear to be unreliable (Gaetz 2004). Furthermore, with
piled-up sources it becomes difficult to obtain the correct PSF
normalization.

In order to address these issues, we chose to develop a representation
of the PSF by utilizing all {\it Chandra}/HETG observations of PKS
2155-304 and Her X-1. These two sources are both out of the plane of
the Galaxy and have interstellar hydrogen column densities of $N_{\rm
  H} \la 10^{20}$ cm$^{-2}$ along the line of sight. Presumably, this
also means that the amount of dust along the line of sight is nearly
negligible, meaning the radial profiles of the zeroth-order images are
almost purely due to the PSF, with only a very small contribution due
to dust-scattering. We use the HETG observations of these sources for
two reasons: (1) The grating allows for flux and spectral
measurements, making it possible to accurately normalize the PSF as a
function of energy; and (2) the use of the grating will provide
consistent analysis when we apply the resulting PSF parameter values
to our observations of Cen X-3 (in case the diffraction of photons
from the HETG modifies the PSF). The drawback to this approach,
however, is that grating reduces zeroth-order effective area and the
statistical quality of the images. To try to overcome the problem of
limited statistics, we co-added the zeroth-order images and
first-order MEG/HEG spectra from five observations of Her X-1 (ObsIDs
2749, 3821, 3822, 6149, \& 6150) and ten observations of PKS 2155-304
(ObsIDs 1014, 1705, 3167, 3706, 3708, 5173, 6926, 7291, 8380, \&
8436), for a total exposure time of 373 ks.

\begin{deluxetable}{clcc} 
\tablenum{1}
\tabletypesize{\scriptsize}
\tablecolumns{4}
\tablewidth{0pt}
\tablecaption{\sc{Chandra/HETG PSF Parametrization}} 
\tablehead{
\colhead{Energy Range} &
\colhead{Normalization\tablenotemark{a}} &
\colhead{} & 
\colhead{} \\
\colhead{(keV)} &
\colhead{($10^{-5}~{\rm arcsec}^{-2}$)} & 
\colhead{$\Gamma$} &
\colhead{$\chi^2_{\nu}$}
}
\startdata       
1.0--1.2 & 9.9 $\pm$ 0.3 (3.6\%)  &  2.62 $\pm$    0.04 &   1.40 \\
1.2--1.4 & 9.1 $\pm$ 0.4 (4.8\%)  &  2.51 $\pm$    0.05 &   0.99 \\
1.4--1.6 & 7.8 $\pm$ 0.5 (6.4\%)  &  2.48 $\pm$    0.07 &   1.24 \\
1.6--1.8 & 7.9 $\pm$ 0.6 (8.4\%)  &  2.30 $\pm$    0.09 &   1.92 \\
1.8--2.0 & 8.7 $\pm$ 0.5 (5.9\%)  &  2.48 $\pm$    0.07 &   1.21 \\
2.0--2.2 & 10.4 $\pm$ 0.6 (6.0\%)  &  2.52 $\pm$    0.07 &   1.44 \\
2.2--2.4 & 9.0 $\pm$ 0.5 (6.8\%)  &  2.44 $\pm$    0.07 &   1.20 \\
2.4--2.6 & 7.7 $\pm$ 0.7 (10.4\%)  &  2.36 $\pm$    0.12 &   2.20 \\
2.6--2.8 & 7.8 $\pm$ 0.6 (8.9\%)  &  2.12 $\pm$    0.09 &   1.60 \\
2.8--3.0 & 8.7 $\pm$ 0.7 (8.8\%)  &  2.20 $\pm$    0.10 &   1.21 \\
3.0--3.2 & 9.6 $\pm$ 0.8 (9.7\%)  &  2.24 $\pm$    0.10 &   0.95 \\
3.2--3.4 & 9.6 $\pm$ 0.8 (8.9\%)  &  2.37 $\pm$    0.10 &   1.12 \\
3.4--3.6 & 10.0 $\pm$ 0.8 (8.7\%) &  2.13 $\pm$    0.09 &   1.48 \\ 
3.6--3.8 & 9.6 $\pm$ 0.8 (9.3\%)  &  2.03 $\pm$    0.09 &   1.23 \\
3.8--4.0 & 10.5 $\pm$ 0.7 (7.4\%) &  1.97 $\pm$    0.07 &   0.73 \\
4.0--4.2 & 10.8 $\pm$ 0.8 (7.8\%) &  2.09 $\pm$    0.07 &   1.16 \\
4.2--4.6 & 11.8 $\pm$ 0.8 (7.2\%) &  2.00 $\pm$    0.07 &   0.97 \\
4.4--4.8 & 11.7 $\pm$ 0.8 (7.7\%) &  2.04 $\pm$    0.07 &   1.29 \\
4.6--4.8 & 11.6 $\pm$ 0.8 (7.3\%) &  1.92 $\pm$    0.06 &   1.07 \\
4.8--5.0 & 12.1 $\pm$ 0.8 (7.4\%) &  1.98 $\pm$    0.06 &   1.83 \\
    
\enddata
\tablecomments{PSF approximated by co-adding HETG observations of Her X-1 (ObsIDs 2749, 3821, 3822, 6149, \& 6150) and PKS 2155-304 (ObsIDs 1014, 1705, 3167, 3706, 3708, 5173, 6926, 7291, 8380, \& 8436). The PSF can be approximated by the function $S_{\rm PSF} (E,\theta) = A_{5}(E)\left(\theta/5\arcsec\right)^{-\Gamma(E)}$, where $A_{5}(E)$ is the normalization 5\arcsec~from the point source. Parameters apply from 3\arcsec~to 100\arcsec~from the central point source; an exponential cut-off is required for analysis of larger off-axis angles (Gaetz 2004).}
\tablenotetext{a}{$A_{5}(E)$: Fraction of the point source flux comprising the PSF 5\arcsec~from the central source in units of $10^{-5}~{\rm arcsec}^{-2}$. The fit parameters $A_{5}(E)$ and $\Gamma$ are correlated, so the uncertainty in $A_{5}(E)$ does not precisely correspond to the uncertainty in the integrated PSF flux ($\int 2 \pi \theta S_{\rm PSF} (E,\theta) d \theta$). The percentage in parenthesis more accurately represents the fractional uncertainty in the integrated PSF flux.}
\end{deluxetable}

For each observation, twenty different radial profiles were extracted
in 200 eV energy bands from 1--5 keV using 40 logarithmically-spaced
bins from 0\farcs492 (1 pixel) to 110\arcsec, with spatial filters
applied to the readout transfer streak and the MEG and HEG diffraction
pattern (offset by -4\fdg725 and 5\fdg235 from a line perpendicular to
the readout) using rectangular regions 10 pixels wide. The radial
profiles from each observation were then summed together and scaled to
the total exposure time, resulting in images with units of
[counts/s/pixel].

To account for differences in the detection efficiencies between the
individual observations, separate instrument maps were created for
each observation and energy band. Traditional instrument maps account
for varying effective area and photon detection efficiencies across the
CCD chip, but for PSF analysis it is only appropriate to use the
effective area at the source position (Gaetz 2004). The aspect
histograms of each observation were then used to project the
instrument maps onto the plane of the sky, resulting in effective area
maps accounting for bad rows on the CCD and the dither of the
telescope. The dithered instrument maps, which showed deviations in
the effective area at the source position of a few percent from
observation to observation, were averaged by weighting each individual
map by the number of counts in each energy and angular bin from the
corresponding observation, correcting for the area of the spatial
filters that were applied to the images. Radial profiles of the
exposures maps were then created for the same annular regions as the
profiles of the zeroth-order images, resulting in maps with units of
[cm$^{2}\times {\rm counts}/{\rm photons}$]. The final
exposure-corrected radial profiles were then created by dividing the
summed counts images by the averaged exposure map in each energy band.
Using the summed spectrum, fluxes in each energy band were measured in
units of [photons/cm$^{2}$/s]. By dividing the exposure-corrected
images by the fluxes in each energy band, we obtained normalized PSF
profiles in 200 eV bands.

To each PSF profile (with background subtracted), we fit the function 
\begin{equation}
S_{\rm PSF} (E,\theta) = A_{5}(E)\left(\frac{\theta}{5\arcsec}\right)^{-\Gamma(E)},
\end{equation}
where $A_{5}(E)$ is the normalization 5\arcsec~from the central point
source in units of flux fraction per square arcsecond. When modelling
the PSF at larger off-axis angles (e.g., $\theta > 100\arcsec$) it is
necessary to include an exponential cut-off in the fitting function to
account for the diffuse PSF wings (Gaetz 2004), but at smaller angles
($\theta < 100\arcsec$) a simple power-law representation appears to
be adequate. The final product that we obtained through this analysis
is the energy- and angle-dependent fraction of the source flux
comprising the PSF, as shown in Table 1. These results can be safely
applied to {\it Chandra}/HETG observations 3\arcsec~to 100\arcsec~from
the central source.

\end{document}